\documentclass[twoside]{phstyle}

\usepackage[T1]{fontenc}
\usepackage{microtype}
\DisableLigatures[f]{encoding = *, family = * }

\usepackage{upref,amsmath,amssymb,amsthm,mathrsfs,epsf,amsfonts}
\usepackage{graphicx,subfigure}
\usepackage{makeidx}
\usepackage{hyperref}
\usepackage{paralist}
\usepackage{multirow}
\usepackage{float}
\usepackage{url}
\usepackage{rotating}
\usepackage{epstopdf}
\usepackage{tablefootnote}
\usepackage{xfrac}
\usepackage{paralist}
\usepackage{caption}

\theoremstyle{plain}

\theoremstyle{definition}
\newtheorem{definition}{Definition}[section]
\theoremstyle{example}
\newtheorem{example}{Example}[section]

\graphicspath{{fig/}{../fig/}{./}}

\makeindex
\begin{document}

\pagenumbering{roman}             
\thispagestyle{empty}

\begin{centering}
\begin{figure}[h!bt]
\centerline{\includegraphics[scale=.25]{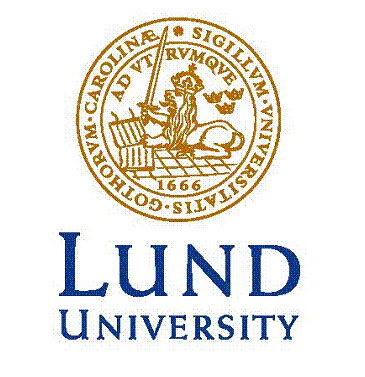}} 
\end{figure}
\rule{5in}{.05in}\\
{\Huge \bf  Spatial methods and their applications to environmental and climate data
\\ \rule{5in}{.05in} \\ \vspace{.2in}}
\begin{figure}[h!bt]
\centerline{\includegraphics[width=0.9\textwidth ,height =0.3\textheight]{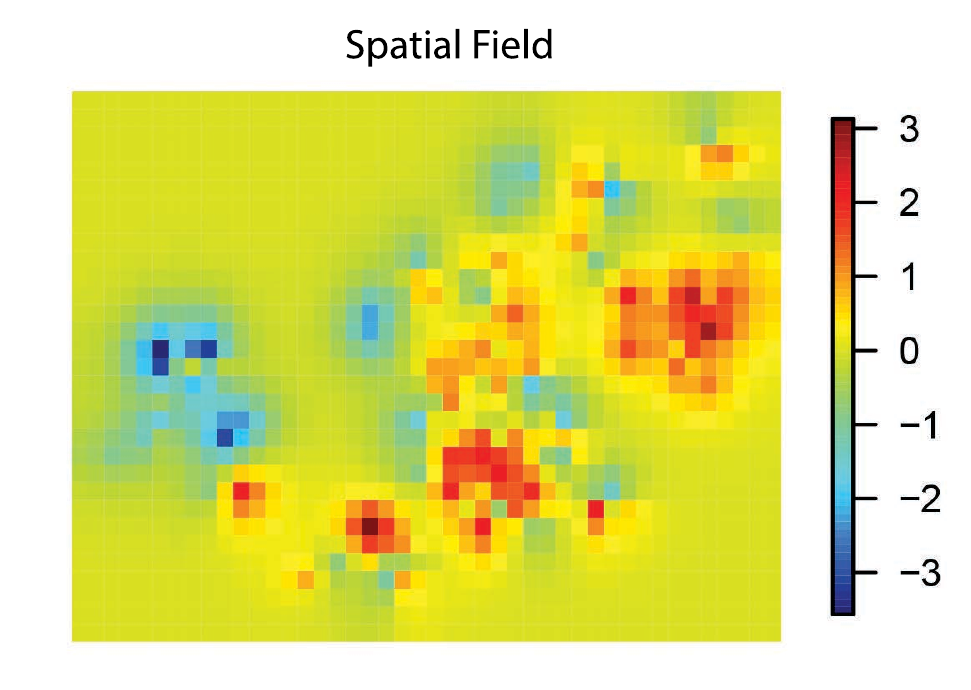}} 
\end{figure}
\vspace{.09in} \vspace{.10in}
\LARGE {\bf Behnaz Pirzamanbin} \\
\vspace{.09in} \vspace{.10in}

\large Department of Mathematical Statistics\\ Lund University, Sweden\\ 2013 \\
\end{centering}
\clearpage

\setcounter{chapter}{0}
\setcounter{page}{1}              

\begin{abstract}
Environmental and climate processes are often distributed over large space-time domains. Their complexity and the amount of available data make modelling and analysis a challenging task. Statistical modelling of environment and climate data can have several different motivations including interpretation or characterisation of the data. Results from statistical analysis are often used as a integral part of larger environmental studies. 

Spatial statistics is an active and modern statistical field, concerned with the quantitative analysis of spatial data; their dependencies and uncertainties. Spatio-temporal statistics extends spatial statistics through the addition of time to the, two or three, spatial dimensions.

The focus of this introductory paper is to provide an overview of spatial methods and their application to environmental and climate data. This paper also gives an overview of several important topics including large data sets and non-stationary covariance structures. Further, it is discussed how Bayesian hierarchical models can provide a flexible way of constructing models. Hierarchical models may seem to be a good solution, but they have challenges of their own such as, parameter estimation.

Finally, the application of spatio-temporal models to the LANDCLIM data (LAND cover - CLIMate interactions in NW Europe during the Holocene) will be discussed. 

\end{abstract}                 

\setcounter{chapter}{0}
\setcounter{page}{1} 
\tableofcontents                  
\parskip 0in                      

\pagenumbering{arabic}            

\chapter{Introduction}\label{intro}
Spatial statistics is an active and modern statistical field, concerned with the quantitative analysis of spatial data, their dependencies and uncertainties. One of the main properties of spatial statistics is the handling of correlated data; i.e allowing for observations that are close to each other, to be more similar than observations that are far apart.

One of the earliest work including spatial considerations was R.A Fisher's  development of design-based inference for agricultural field experiments ($1919$ to $1933$, published in $1966$). Fisher noted that field plots, rectangular units in the field, close to each other were more similar than the plots further apart, violating the assumption of independent data. To account for the increasing dependence, Fisher suggested the use of blocking of the plots, a form of covariate adjustment. Hence, the larger blocks of plots where approximately independent, and spatial variation, if it exists, was constant within blocks. An alternative strategy suggested by Papadakis \cite{papadakis1937methode} is to adjust the plot yields to take account of the average yield in neighbouring plots instead of the overall yield. This strategy has a close relation to Markov random fields as pointed out by Cox et al. \cite{cox1977role} and Bartlett  \cite{bartlett1978nearest} (see further \cite{history-Handbook}). Further development of this concept lead to the general case of Gaussian Markov Random Fields which are discussed in Section \ref{sec:GMRF}.

Other important work in spatial statistics was done by Krige \cite{krige1953statistical} and Math\'{e}ron \cite{matheron1963geostatistics}. Their work focused on the characterization and prediction of spatial data, leading to Geostatistics. Another statistician who has done influential work in modelling spatial dependencies is Bertil Mat\'{e}rn, a Swedish forestry statistician. Mat\'{e}rn's doctoral dissertation is a remarkable work in spatial statistics, having one of the highest number of citations in the field \cite{matern1960spatial}. He introduced the Mat\'{e}rn family of covariance functions which is one of the most popular models for many geostatistical applications \cite{history-Handbook}. 

Spatial statistics can be applied to data from many different fields, including climate and environmental data. The development of spatio-temporal statistics during the last century has aided scientists in solving numerous environmental and climate problems. 
The collaboration between statistician and environmental scientists has also led to important developments in spatial statistics, from the first work of Fisher \cite{fisher1935design}, handling data from agricultural experiments, to modern day methods, handling the very large data sets that arise in climatological applications.
For example, environmental questions regarding aerosol forcing---the global distribution of aerosols, the transport of aerosols, and differences between satellite observations and global-climate-model outputs---were a motivation for equipping satellites to collect global data. These observations are noisy and contain missing values, Shi and Cressie \cite{fixed-ENV:ENV864} used spatial statistics to reconstruct the missing values and denoise the data. In Paleoecology, scientists have questions about past forest composition: how relative vegetation abundances change over time, if forest compositions constantly shifting, how forest changed in response to past climate shift and how humans affect forests in comparison to natural forest change. A recent study by Paciorek and McLachlan \cite{paciorek2009mapping} on US data used a multivariate spatio-temporal process to model forest composition for past time periods, based on observations from the colonial era (1635-1800) and 20th centuries. The model of \cite{paciorek2009mapping} estimates the spatial distribution of the relative vegetation abundances and gives uncertainties for these estimates.

Throughout this paper more environmental problems are given as examples in occurrence with the statistical methods that used to solve them.

This paper will provide an overview of spatial statistics and its applications. The first section provides some basic concepts in probability. Section \ref{sec:spatial} gives an introduction to spatial statistical modelling, covering parameter estimation and prediction. In Section \ref{sec:spatio-temp} spatio-temporal model are introduced. Section \ref{sec:non-stat} introduces one of the challenges in spatial statistics, non-stationary covariance structures; some common solutions to the problem are also discussed. Section \ref{sec:largedata} discusses the issue of large spatial data sets. Finally, Section \ref{sec:application} discusses how spatio-temporal models can be applied to the LANDCLIM data (LAND cover - CLIMate interactions in NW Europe during Holocene), one of the projects in strategy research areas MERGE (ModElling the Regional and Global Earth system) with the focus on \textbf{land-cover/vegetation}.
 
\section{Basic concepts}
\label{sec:one}

\subsection{Properties of random variable}

Given a continuous random variable with density $ f(x) $ the expectation is defined as $E(x) = \int xf(x) dx$ ( or $E(x) = \sum x\textbf{p}(x)$ in discrete form) and the variance is $V(x) = E(\left[x-E(x)\right]^2) = E(x^2)-E(x)^2 .$
The covariance of a bi-variate random variable is defined as
\begin{equation}
C(x,y) = E\left(\left[x-E(x)\right]\left[y-E(y)\right]\right)  = E(xy) - E(x)E(y)=C(y,x). 
\label{eq:cov}
\end{equation}
And for multivariate random variables the covariance matrix is constructed as $\Sigma_{ij}=C(x_i,x_j)$, note that $\Sigma_{ii}=C(x_i,x_i)=V(x_i)$.
The covariance matrix has three properties; it is \begin{inparaenum}[i\upshape)]\item Square \item Symmetric, $ \Sigma = \Sigma^T $ and \item Positive definite, $ a^T\Sigma a > 0 $ if $ a\neq 0 $.\end{inparaenum}~The first two properties follow trivially from \eqref{eq:cov}, and are easy to verify. The following proof shows why the covariance matrix needs to be positive definite.

\textbf{Proof}\:(iii)~Consider the covariance in matrix form,
\begin{eqnarray}
\begin{split}
 a^T\Sigma a & = E[a^T(\mathbf{X}-\mathbf{\mu})(\mathbf{X}-\mathbf{\mu})^Ta]\\
             & = E[\big(a^T(\mathbf{X}-\mathbf{\mu})\big)^2]=V\big(a^T\mathbf{X}\big)\\
             & > 0\qquad  \text{if} \; a\neq 0 \quad \text{and} \quad V(\mathbf{X})\succ 0.
\end{split}
\label{eq:proof}
\end{eqnarray}
From the proof above one can see that the variance of linear combinations(e.g. a mean) of random variables is given by a quadratic-form involving the covariance matrix. Additionally, verifying that a matrix is positive definite, i.e. a valid covariance, is hard since $ a^T\Sigma a > 0 $ must hold for {\em all} $ a\neq 0 $; this will present a challenge when constructing covariance matrices.

\subsection{Linear regression}
\label{sec:reg}
A statistical model is a family of probability distributions, $p$, which one assumes that a particular data set, $y$, is sampled from. For a parametric model, there are {\em unknown} parameters, $\theta$, which control the distribution, $p$. Linear regression \cite{rawlings1998regression} is an example of a simple statistical model. 
\begin{definition}[Linear regression]
Let $Y=(y_1,y_2,\dotsc,y_n)^T$ be a vector of observations and $X=(X_1,X_2,\dotsc,X_p)$ be explanatory variables that are fixed and known, then the linear model is constructed as 
\[y_i = \sum_p{X_{i,p}\beta_p} + e_i \qquad e_i \in \mathcal{N}(0,\sigma^2)\]  
or in matrix form,
\[\textbf{Y} = \textbf{X}{\boldsymbol\beta}+ \textbf{e} \qquad \textbf{e} \in \mathcal{N}(0,\mathbb{I}\sigma^2).\]
In linear regression one assumes that the observations $\mathbf{Y}$ follow a Gaussian distribution with a mean that depends {\em linearly} on a set of known explanatory variables $\mathbf{X}$. The unknown parameters are: how much of each explanatory variable should be used in the mean $\beta$,and how large the variability, $\sigma$, around the mean is.
\label{def:reg}
\end{definition}
An estimate of $\beta$ is 
\[
\hat{{\boldsymbol\beta}} = ( \mathbf{X}^T\mathbf{X} ) ^{-1}  \mathbf{X}^T\mathbf{Y}.
\]
Moreover, under the model assumptions, the residuals $\hat{\textbf{e}}=\textbf{Y} - \textbf{X}\hat{{\boldsymbol\beta}}$, have the following properties,
\begin{inparaenum}[1\upshape)]
\item They are independent,
\item Normally distributed,
\item Have equal variance.
\end{inparaenum}

In order to choose a better regression model, typically a model with optimal number of explanatory variables, one can perform different statistical tests. One way of doing this is to define a criteria for the best model. Two common criteria are Akaike's information criterion (AIC)~\cite{akaike1974new},
\[
AIC(p+1)= -2\log L(\hat{\beta},\hat{\sigma})+2(p+1),
\]
and Schwartz's Bayesian information criterion (BIC)~\cite{schwarz1978estimating},
\[
BIC(p+1)= -2\log L(\hat{\beta},\hat{\sigma})+\log n(p+1).
\]
In general the second term in both AIC and BIC increases with $p$, however larger $p$ gives smaller $L(\hat{\theta})$, i.e. likelihood at $\hat{\theta}$, thus AIC and BIC attempts to balance between model size and residual error. For AIC the model size penalty $2(p+1)$ does not depend on number of observation, $n$, which, especially for large $n$, can be a problem. The model size penalty is adjusted in BIC with $\log n$, hence BIC typically give smaller models.

\subsection{Parameter estimation}
\label{sec:LSML}
A common problem in statistics is to estimate parameters of a model. Given a model $ p(y; \theta) $ with parameter(s) $\theta $, two common methods for estimation of the parameters are \cite{lehmann2005LSML}:
\begin{enumerate}
\item Least squares (LS)
\item Maximum likelihood (ML)
\end{enumerate}

In least squares one tries to find the parameter(s) that minimizes the sum of square errors between the observation and expected value given parameter(s),
\[
\hat{\theta} = \underset{\theta}{\operatorname{argmin}}\sum_{i=1}^n \left(  y_i-E(y_i;\theta)\right)   ^2 
\]
and in the case of maximum likelihood, assuming that observations are independent the likelihood is 
\[
L(\theta|y_1,\dotsc,y_n) = p(y_1,y_2,\dotsc; \theta)\underset{indp}{\operatorname{=}}\prod_{i=1}^n p(y_i;\theta)
\]
hence, one tries to find parameter(s) that maximize the likelihood function,
\[
\hat{\theta} = \underset{\theta}{\operatorname{argmax}}~L(\theta|y_1,\dotsc,y_n).
\]
\\                
\chapter{Spatial Statistics}
\label{sec:spatial}

\section{Theory of Gaussian fields}
\label{sec:Theory}
A statistical model describes and analyses randomness in a set of data. For spatial data, one possible model is a random, or stochastic, field.
\begin{definition}[Stochastic field]
A random or stochastic field, $X(\mathbf{u}), \: u\in \mathbb{D} \subseteq \mathbb{R}^d$ is a random function specified by its finite-dimensional joint distributions 
\[
F_{u_1,\dotsc ,u_n}(x_1,\dotsc ,x_n)= \mathbf{P}(X(u_1)\leq x_1, \dotsc , X(u_n)\leq x_n)
\]
for all finite n and all collection ${u_1, \dotsc, u_n}$ of locations in $\mathbb{D}$.
\end{definition}
Consider a field $x$ defined in two dimensions, the field is called a Gaussian random field if any subset of points in the field, ${x(u_1),\cdots,x(u_n)}$, are jointly multivariate Gaussian, i.e. $\mathbf{x} \in \mathcal{N}(\mu,\Sigma)$, with density given by
\begin{equation}
p(\mathbf{x})= \dfrac{1}{(2\pi)^{N/2}|\Sigma|^{1/2}}exp\left( -\frac{1}{2}(\mathbf{x}-\mu)^T\Sigma^{-1}(\mathbf{x}-\mu)\right) .
\label{eq:normdensity}
\end{equation}
Moreover, the field has a expectation function $\mu_x(\mathbf{u})=E(\mathbf{x}(\mathbf{u}))$ and covariance function
\[
r_x(\mathbf{u},\mathbf{v})= C(\mathbf{x}(\mathbf{u}),\mathbf{x}(\mathbf{v})).
\]
The stochastic field can further be stationary and/or isotropic.

\begin{definition}[$2^{nd}$ order/weak stationary]
A field is said to be $2^{nd}$ order stationary if expectation and covariance are unchanged under translation. 
\begin{itemize}
\item $\mu_x(\mathbf{u})= \mu_x(\mathbf{u}+h) = constant.$
\item $r_x(\mathbf{u},\mathbf{v})= r_x(\mathbf{u}+h,\mathbf{v}+h)\;\Longrightarrow  \; r_x(0,\mathbf{v-u})= r_x(h)$.
\end{itemize}
A stationary field is sometimes said to be homogeneous.
\end{definition}

\begin{definition}[Isotropic]
If a stationary covariance depends only on the distance between the points and not on the direction, $r(h)=r(\Vert h \Vert)$, the field is said to be isotropic; otherwise it is anisotropic. 
\end{definition}

An alternative to the covariance function is the semi-variogram. Semi-variograms can be defined even if the covariance function does not exit; e.g. if the field lacks finite expectation $E(x)\nless \infty$.
 
\begin{definition}[Semi-variogram]
\label{def:semiV}
A semi-variogram for a stationary and isotropic field is defined as
\begin{equation}
\gamma(\Vert h\Vert)= \frac{1}{2}V(\mathbf{x}(\mathbf{u}+h)-\mathbf{x}(\mathbf{u}))= r(0)-r(\Vert h\Vert)
\label{eq:semi}
\end{equation}
Estimation of semi-variograms is more robust to miss-specification of the mean than estimation of covariance function. This, since the mean cancels in the subtraction of  $\mathbf{x}(\mathbf{u}+h)$ and $\mathbf{x}(\mathbf{u})$. If the field has constant mean then \eqref{eq:semi} becomes
\[
\gamma(\Vert h\Vert)= \frac{1}{2}E\big[(\mathbf{x}(\mathbf{u}+h)-\mathbf{x}(\mathbf{u}))^2\big]
\]
\end{definition}


In Table~\ref{covfun} some common covariance functions and corresponding semi-variogram are shown.

\begin{table}[H]
\small
\begin{tabular}{l|l|l}

Name & Covariance, $r(h)$ & Semi-variogram, $\gamma(h)$\\
\hline
\hline
&\\[0.1cm]
Mat\'{e}rn & $\sigma^2\dfrac{(\sfrac{h}{\rho})^\nu\mathbf{K}_\nu(\sfrac{h}{\rho})}{\Gamma(\nu)2^{\nu-1}}$& $\sigma^2\left(1-\dfrac{(\sfrac{h}{\rho})^\nu\mathbf{K}_\nu(\sfrac{h}{\rho})}{\Gamma(\nu)2^{\nu-1}}\right)$\\[0.5cm]

Exponential\tablefootnote{Mat\'{e}rn with $\nu=1/2$.}& $\sigma^2exp(-\sfrac{h}{\rho})$ & $\sigma^2\big(1-exp(-\sfrac{h}{\rho})\big)$\\[0.5cm]

Gaussian\tablefootnote{Mat\'{e}rn with $\nu \rightarrow \infty $.}& $\sigma^2exp(-(\sfrac{h}{\rho})^2)$ & $\sigma^2\big(1-exp(-(\sfrac{h}{\rho})^2)\big)$\\[0.5cm]

Spherical & $\left\{ 
  \begin{array}{l}
    \sigma^2\left(1-1.5(\frac{h}{\rho})+0.5(\frac{h}{\rho})^3\right)\\
    0\\
  \end{array} \right.$ & 
  $\left\{ 
  \begin{array}{l l}
    \sigma^2\left(1.5(\frac{h}{\rho})-0.5(\frac{h}{\rho})^3\right) & h<\rho\\
    \sigma^2 &  \text{o.w}\\
  \end{array} \right.$\\
 
\end{tabular}
\caption{Different covariance functions and semi-variograms, in general $\sigma^2$ denotes the variance of the field and $\rho$ is the range parameter.} 
\label{covfun}

\end{table}

\section{The basic model}
Geostatistical measurements are often made with some noise or error in measurements, therefore any model describing the data has to account for the noise. Generally, the simplest model consists of a latent Gaussian field $\textbf{X}\in \mathcal{N}(\mu,\Sigma)$ at some locations $\{u_i\}_{i=1}^n$ observed with additive noise,
\[
y_i=x(u_i)+\varepsilon_i \,, \qquad \varepsilon_i \in \mathcal{N}(0,\sigma_\varepsilon^2),
\]
or in matrix form
\begin{equation}
\mathbf{Y}= A\mathbf{X}+\varepsilon\,, \qquad \varepsilon \in \mathcal{N}(0,\mathbb{I}\sigma_\varepsilon^2)
\label{eq:yAx}
\end{equation}
where the matrix $A$ is a sparse observation matrix that extracts the appropriate
elements from $\textbf{X}$. The joint distribution of $\textbf{X}$ and $\textbf{Y}$ is 
\begin{equation}
\textbf{Z} =
 \begin{bmatrix}
 \textbf{X}\\
 \textbf{Y}
\end{bmatrix}
\in \mathcal{N}
\left( 
\begin{bmatrix}
 \mu_{\mathbf{x}}\\
 \mu_{\mathbf{y}}
\end{bmatrix}
,
\begin{bmatrix}
 \Sigma_{\mathbf{x}\mathbf{x}}& \Sigma_{\mathbf{x}\mathbf{y}} \\
 \Sigma_{\mathbf{y}\mathbf{x}}& \Sigma_{\mathbf{y}\mathbf{y}} 
\end{bmatrix}
\right)
=
\left( 
\underbrace{\begin{bmatrix}
 \mu\\
 A\mu
\end{bmatrix}}_{\mu_\textbf{z}}
,
\underbrace{\begin{bmatrix}
 \Sigma& \Sigma A^T \\
 A\Sigma& A\Sigma A^T+\mathbb{I}\sigma_\varepsilon^2 
\end{bmatrix}}_{\Sigma_\textbf{z}}
\right)
\label{eq:condition}
\end{equation}
or
\[
\mathbf{Z} \in \mathcal{N} \left( \mu_\textbf{z}(\theta), \Sigma_\textbf{z}(\theta) \right). 
\]
Typically the latent field is decomposed into a deterministic mean and a stationary, mean zero field,
\begin{equation}
\textbf{X}=\mu+\eta,, \qquad \eta \in \mathcal{N}(0,\Sigma(\theta)).
\label{eq:xmunu}
\end{equation}
The mean is often modelled as a regression, $\mu=\textbf{B}\beta$.

A standard problem in spatial statistics is to reconstruct the latent field, $\textbf{X}$ given observations $\textbf{Y}$. For known parameters this is described in Section \ref{Pred}, however parameters are often unknown and need to be estimated, see Section \ref{sec:PE} for more details.

\section{Prediction}
\label{Pred}
The most famous method for reconstructing the field is kriging. The method was developed by Math\'{e}ron \cite{matheron1963geostatistics} from Krige \cite{krige1953statistical} master's thesis in geostatistics. It was mostly focus on spatial dependency and predicting values of field over a spatial region.

Assuming a \textbf{known covariance}, ${\boldsymbol\Sigma}$, a regression formulation of the expectation, $\mu = B\beta$, and given the observation at some location, $y(u_i), \: i=1\cdots n$, the aim is to predict values, i.e.~reconstruct the field, at unobserved location, $x(u_s)$. The reconstructions should be: \begin{inparaenum}[1\upshape)]
\item Linear in the observations, $ \hat{x}_{s}= \sum_k \lambda_k y(u_k)$;
\item Unbiased, $E(\hat{x}_{s}) = E(x_{s})$; and have
\item Minimum prediction variance; $\min_{\lambda} V(\hat{x}_{s}-x_{s})$.
\end{inparaenum}
Traditionally Kriging has been divided into three different cases: \begin{inparaenum}[1\upshape)]
\item Simple Kriging: $\mu$ is known, 
\item Ordinary Kriging: $\mu$ is unknown, but constant,
\item Universal Kriging: $\mu = B\beta$, with $\beta$ unknown. (Possibly, estimation of unknown $\Sigma$.)
\end{inparaenum}
Note that $\mu=\mathbb{I}\beta$ gives an unknown constant making $2)$ a special case of $3)$.

For a Gaussian process the best linear unbiased predictor is the conditional expectation \cite{best-prediction-Handbook},
\begin{eqnarray}
\begin{split}
E(X|Y)&= B_x\hat{\beta}+\Sigma_{xy}\Sigma_{yy}^{-1}(Y-B_y\hat{\beta})\\
&= \big(B_x-\Sigma_{xy}\Sigma_{yy}^{-1}B_y\big)\hat{\beta}+\Sigma_{xy}\Sigma_{yy}^{-1}Y
\end{split}
\label{eq:estimation}
\end{eqnarray}
where 
\[
\hat{\beta} =\left(B_y^T\Sigma_{yy}^{-1}B_y\right) ^{-1}B_y^T\Sigma_{yy}^{-1}Y
\]
and $B_x$ is the covariate of the regression model used to estimate $\mu$ and $B_y = A*B_x$.
The prediction uncertainty for simple kriging is 
\begin{equation}
V(X|Y) = \Sigma_{xx}-\Sigma_{xy}\Sigma_{yy}^{-1}\Sigma_{yx}^T
\label{eq:predv}
\end{equation}
and for the ordinary and universal kriging the $\beta$ estimator's uncertainty is added to \eqref{eq:predv},
\begin{eqnarray*}
V(X|Y) & = & \Sigma_{xx}-\Sigma_{xy}\Sigma_{yy}^{-1}\Sigma_{yx}\\
& + &
\left(B_x^T-B_y^T\Sigma_{yy}^{-1}\Sigma_{yx}\right)^T \underbrace{\left(B_y^T\Sigma_{yy}^{-1}B_y\right)}_{V(\hat{\beta}|Y)}
\left(B_x^T-B_y^T\Sigma_{yy}^{-1}\Sigma_{yx}\right).
\end{eqnarray*}

\begin{example}[Temperature map using Universal Kriging]
This example shows the estimated temperature for winter in U.S using 250 locations. The estimated values are predicted using universal kriging. In Figure \ref{fig:Avg_temp_uni_krig}, observed locations, prediction and standard error (SE) are shown.
\begin{figure}[H]
\centerline{\includegraphics[scale=0.5]{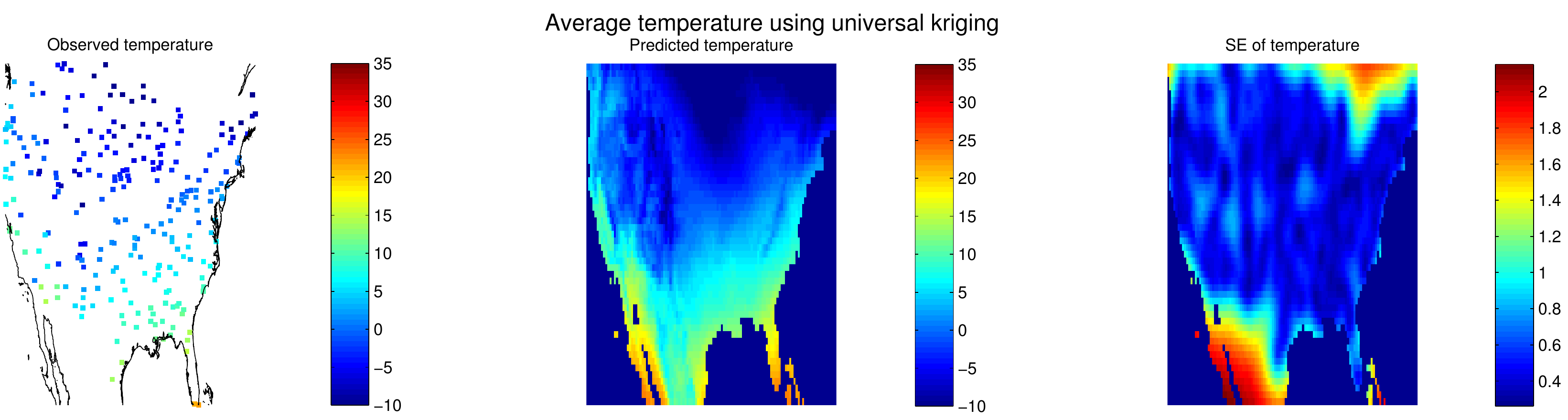} }
 \caption{Estimated temperature for winter in U.S using universal kriging from $250$ locations.}
 \label{fig:Avg_temp_uni_krig} 
 \end{figure} 
\end{example} 
 
\section{Parameter estimation}
\label{sec:PE}

The reconstruction assumes a known covariance matrix $\Sigma$. However, in practice $\Sigma$ often has to be estimated from data. Properties of $\Sigma$, such as symetry and positive definite, put constrains on the estimation. To solve this problem one often assumes that the covariance matrix is defined by a parametric family of covariance functions. This assumption reduces the problem to estimate the parameters of the covaraiance function. Uncertainties in the estimated parameters can be accounted for by using Markov Chain Monte Carlo (MCMC) methods \cite{hastings1970monte}\label{mcmc} or numerical integration (INLA) \cite{rue2009approximate}.

In order to estimate the covariance function using a non-parametric approach, one might use the semi-variogram defined in Definition \ref{def:semiV}. One can estimate parameters of covariance function using the two common methods explained in Section \ref{sec:LSML}; LS or ML. In LS, an estimated mean is subtracted from the observation, $\mathbf{w}=\mathbf{Y}-\mathbf{B}\hat{\beta}$, and the semi-variogram,
\[ 
\hat{\gamma}(\Vert \mathbf{u_i}-\mathbf{u_j} \Vert) = \frac{1}{2}E\left( \mathbf{w}(\mathbf{u_i})-\mathbf{w}(\mathbf{u_j})\right) ^2,
\]
gives one estimate for every pair of observations. The estimates, $\hat{\gamma}$, are often binned, i.e.~grouped into blocks with similar distances $(u_i-u_j)$. The estimates are then averaged within each block. In order to handle the semi-definite restriction one may choose a parametric semi-variogram, e.g.~one of those in Table \ref{covfun}, and pick parameters, $\theta$ such that $\hat{\gamma}$ and $\gamma(h;\theta)$ are as close as possible, that is 
\begin{equation}
\hat{\theta}= \underset{\theta}{\operatorname{argmin}}~\sum \big(\gamma(h;\theta)-\hat{\gamma}\big)^2.
\label{eq:LS-min}
\end{equation}
Parameter estimation in LS can be sensitive to the number of bins. 

In ML one assumes that the observations, $\mathbf{Y}$, form a Gaussian field with mean given by a regression model, $\mu=\mathbf{B}\beta$ and covariance given by a covariance function, $r(h;\theta)$ with unknown parameters, $\theta$,
\[
\mathbf{Y} \in \mathcal{N}\big(\mathbf{B}\beta,\Sigma(\theta)\big).
\]
The log-likelihood of $\mathbf{Y}$ is 
\[
\ell(\theta,\beta|\mathbf{Y})= \text{constant}-\frac{1}{2}\log\vert \Sigma(\theta)\vert -\frac{1}{2}(\mathbf{Y}-\mathbf{B}\beta)^T\Sigma^{-1}(\theta)(\mathbf{Y}-\mathbf{B}\beta),
\]
and ML-parameter estimation is given by maximizing $\ell$.
\[
(\hat{\theta},\hat{\beta})=\underset{\theta,\beta}{\operatorname{argmax}}~\ell(\theta,\beta|\mathbf{Y}),
\]
The maximization can be done in two steps. First, for any fixed value of parameters $\theta_0$ there is a unique value of $\beta$ that maximize $\ell$. Second, for a given $\beta(\theta_0)$ there is a $\hat{\theta}$ that maximize $\ell(\theta;\mathbf{Y})$. Eliminating $\beta$ from the log-likelihood function in these two steps is called profiling, and $\ell(\theta;\mathbf{Y})$ is called the profile log-likelihood,
\begin{equation}
\ell_{\text{profile}}(\theta;\mathbf{Y})=-\frac{1}{2}\log\vert\mathbf{\Sigma(\theta)}\vert-\frac{1}{2}\mathbf{Y^TP(\theta)Y},
\label{eq:profile}
\end{equation}
and 
\begin{equation}
\hat{\beta}(\theta)= \left(\mathbf{B^T\Sigma^{-1}(\theta)B}\right)^{-1}
\mathbf{B^T\Sigma ^{-1}(\theta)Y},
\label{eq:beta}
\end{equation}
where 
\[
\mathbf{P(\theta)}=\mathbf{\Sigma^{-1}(\theta)-\Sigma^{-1}(\theta)B\left(B^T\Sigma^{-1}(\theta)B\right)^{-1}X^T\Sigma^{-1}(\theta)}.
\]

Even though ML is the most common way of estimating parameters, ML estimators are biased due to the reduction in degrees of freedom caused by the estimation of $\beta$ \cite{harville1977maximum}. Instead one can use the restricted maximum likelihood (REML) which reduces or even eliminates the bias in $\hat{\theta}$. 

REML estimates the parameter by maximizing the log-likelihood function associated with error contrasts. The error contrasts are the ($n-p$) linearly independent combination of observations, $a^TY$ that have expectation zeros for all $\beta$ and $\theta$. This assumption adds an extra constant term to \eqref{eq:profile},
\[
\ell_{REML}(\theta;\mathbf{Y})=-\frac{1}{2}\log\vert\mathbf{\Sigma(\theta)}\vert-\frac{1}{2}\log\vert\mathbf{B^T\Sigma^{-1}(\theta)B}\vert-\frac{1}{2}\mathbf{Y^TP(\theta)Y}
\]
where the ignored additive constant does not depend on the parameters ($\beta, \theta$). A REML estimate of $\theta$ is any values $\tilde{\theta}$, that maximize $\ell_{REML}$. Once the estimator, $\tilde{\theta}$, has been obtained the corresponding estimate of $\beta$ is obtained through \eqref{eq:beta}.
Although, REML decreases the bias it may increase the variance of the estimation \cite{REML-Handbook}.
 
\section{Hierarchical models}
The spatial fields are often used as components of hierarchical models, allowing for a flexible model description and better handling of model uncertainties.
 
A hierarchical model is based on the joint distribution of a collection of random variables as a series of conditional distributions and a marginal distribution. Assume $A, B~\text{and}~C$ are random variables, then one can write the joint distribution as
\[
[A,B,C]=[A|B,C][B|C][C]
\]
where $[C]$ is probability distribution of $C$ and $[B|C]$ is the conditional distribution of $B$ given $C$ and so on.
The simplest hierarchical model consists of three parts:
\begin{enumerate}
\item \textbf{Data model}, $[\text{Data}|\text{Process,Parameters}]$,
\item \textbf{Process model}, $[\text{Process}|\text{Parameters}]$,
\item \textbf{Parameter model}, $[\text{Parameters}]$.
\end{enumerate}
Often one is interested in the distribution of the process and parameters given the data which is called posterior distribution. Using Bayes formula one can write the posterior as a 
\begin{align*}
[\text{Process, Parameters}|\text{Data}]&\propto\\
[\text{Data}|\text{Process},\,&\text{Parameters}][\text{Process}|\text{Parameters}][\text{Parameters}]
\end{align*}

In the spatial context one can specify the hierarchical model terms by
\begin{description}
  \item[Data model:] Describes the distribution of the measurements given the latent process $\mathbf{Y|X},{\boldsymbol\theta} \in \mathcal{N}(\mathbf{B}{\boldsymbol\beta}+\mathbf{AX},\mathbb{I}\sigma^2_\varepsilon)$.
  \item[Latent process:] Describes how the latent variables behave, $\mathbf{X}|{\boldsymbol\theta} \in \mathcal{N}(\mathbf{B}{\boldsymbol\beta},{\boldsymbol\Sigma})$.
  \item[Parameters:] Describes prior knowledge or assumptions regarding the parameters,  $\pi({\boldsymbol\theta})$.
\end{description}
Therefore, the posterior is 
\[
\pi(\mathbf{X},{\boldsymbol\theta}|\mathbf{Y})\propto \pi(\mathbf{Y}|\mathbf{X},{\boldsymbol\theta})\pi(\mathbf{X}|{\boldsymbol\theta})\pi({\boldsymbol\theta})
\]
and the marginal posterior distribution is 
\[
\pi(\mathbf{X|Y}) \propto \int \pi(\mathbf{X|Y},{\boldsymbol\theta})\pi({\boldsymbol\theta}|\mathbf{Y})\mathrm{d}{\boldsymbol\theta}.
\]
Accounting for parameter uncertainties the posterior mean $E(\mathbf{X|Y})$ and posterior variance $V(\mathbf{X|Y})$ provide predictions and predictions uncertainty. 
\\
\chapter{Spatio-temporal statistics}
\label{sec:spatio-temp}
A spatio-temporal process is a process which varies in both space and time. Consider a spatial stochastic processes, $\lbrace X(\mathbf{s}) : \mathbf{s} \in \mathbb{R}^d\rbrace$ then a spatio-temporal processes can be express as
\[
\lbrace X(\mathbf{s,t}) : (\mathbf{s,t}) \in \mathbb{R}^d \times \mathbb{R}  \rbrace.
\]
Hence, $X(s,t)$ is a function of both spatial locations, $\mathbf{s}\in \mathbb{R}^d$, and time, $\mathbf{t} \in \mathbb{R}$. Time can be considered as an additional coordinate, and thus the domain of process becomes $\mathbb{R}^{d+1}=\mathbb{R}^d \times \mathbb{R}$. Often spatio-temporal dependence can be modelled by a spatial process that dynamically changes in time \cite{sptio-tempo-Handbook}. For example, assume a time discrete, spatially continuous process $X_t = \lbrace X(s_1,t), \dotsc ,X(s_n,t)\rbrace$, in the simplest case this process can be modelled as
\begin{eqnarray*}
X_t&=&DX_{t-1}+\nu_t \quad\quad \nu_t  \in \mathcal{N}(0,\Sigma_\nu),\\
Y_t&=&CX_t+\varepsilon_t \quad\qquad \varepsilon_t  \in \mathcal{N}(0,\Sigma_\varepsilon),
\end{eqnarray*}
where $D$ is the state transition matrix, $\Sigma_\nu$ is the covariance matrix of the driving spatial process, $Y_t$ is the observation, $C$ is a sparse observation matrix that extracts the appropriate
elements from $X_t$ and $\Sigma_\varepsilon$ is the covariance of the observations. Note that the covariance matrices, $\Sigma_\varepsilon ~\text{and}~\Sigma_\nu$ can depend on time, providing a highly complicated process. To avoid a too complex spatio-temporal process some simplifying assumptions are necessary. For instance, if a spatio-temporal covariance matrix is separable it can be decomposed into the product of a purely spatial and a purely temporal covariance function. The assumption of separability simplifies the construction of the model and reduces both the number of unknown parameters and the computational time \cite{sptio-tempo-Handbook}.

Spatio-temporal processes have been applied to different environmental problems. For example, Gelfand et al. \cite{spatio-temporal-gelfand2005} used the spatio-temporal methods to model rain fall/ precipitation. Cameletti et al. \cite{spatio-temporal-cameletti2011comparing} and Sampson et al.  \cite{spatio-temporal-Sampson11} applied spatio-temporal methods to model the air quality and pollution.

\chapter{Non-stationary covariance structures}
\label{sec:non-stat}
So far in this paper the covariance functions have been assumed to be stationary. For most environmental processes the covariance structure is non-stationary when considered over large enough spatial scales. However, in most cases a non-stationary model can be seen as a combination of small scale stationary components. A main idea when solving non-stationary problems is to consider small/local spatial regions, i.e.~to assume that the covariance structure is locally stationary. Here, some of these methods are reviewed: \begin{inparaenum}[1\upshape)]
\item \textbf{Process convolution} \cite{higdon1998process},
\item \textbf{Deformation approach} \cite{sampson1992nonparametric}
and \item Allowing for \textbf{covariates} in the covariance structure \cite{schmidt2011considering}.
\end{inparaenum}

\section{Process convolution}
\label{ProCon}
In the process convolution method, the Gaussian random field, $\mathbf{X(u)}$ in $\mathbb{R}^d$ expressed as 
\begin{equation}
\mathbf{X(u)}=\int k(\textbf{s,\;u})w(\textbf{s})d\textbf{s}
\label{eq:Gkernel}
\end{equation}
where $w$ is a Gaussian process and $k$ is a convolution kernel. This can be used to produce non-stationary models by allowing the convolution kernel to depend on location. If the process is stationary then $k(s,u)=k(s-u)$ and the covariance function for $\mathbf{X(u)}$ depends only on $h=u-u'$ and is given by
\[
r(h)=cov(\mathbf{X}(u),\mathbf{X}(u'))=\int k(s-u)k(s-u')ds=\int k(s-h)k(s)ds=k\ast k.
\]
Moreover, the covariance function $r$ and the kernel $k$ are related through Fourier transform such that
\begin{equation}
\mathcal{F}(r)=\mathcal{F}(k)\cdot \mathcal{F}(k)
\label{eq:F}
\end{equation} 

Note that the shape of the kernel determines the shape of the local spatial covariance function. For example, a Gaussian kernel correspond to a Gaussian covariance, and a Mat\'{e}rn kernel leads to a Mat\'{e}rn covariance function. 
In practice \eqref{eq:Gkernel} can be approximated by discretizing the area into intervals, $I_i$ centred at some fixed locations $u_i$'s and writing \eqref{eq:Gkernel} as
\[
\mathbf{X(u)}=\sum_i \int_{s\in I_i} k(u-s)w(s)ds
\]
and using integral approximation
\begin{equation}
\mathbf{X(u)}\approx \sum_{i=1}^m k(u-s_i)w_i
\label{eq:procon}
\end{equation}
where $m$ is the number of intervals on $\mathbb{R}^d$ and $w_i$ are independent zero mean Gaussian variables with variance equal to the area of $I_i$.
Simpson et al. \cite{simpson2010order} showed that approximation \eqref{eq:procon} does not work for general Mat\'{e}rn fields, it only works for fields that have smoothness parameter $\nu_c>d/2$. It can be shown through the Fourier transform relation \eqref{eq:F} that for any value of the smoothness $\nu_k$ in the Mat\'{e}rn kernel, the Mat\'{e}rn covariance, see Table \ref{covfun}, will have a value $\nu_c=2\cdot\nu_k+d/2$, this implies a lower limit on the smoothness of the process, i.e. $\nu_c>d/2$ in order to satisfy $\nu_k>0$. If  $\nu_k\leqslant 0$ the Mat\'{e}rn kernel is singular. Simpson et al. \cite{simpson2010order} solved the problem by modifying \eqref{eq:procon} to a more general and appropriate discretization 
\[
\sum_i \big(\dfrac{1}{|I_i|}\int_{I_i} k(s-u_i)du \big) w_i.
\]
Examples of applying convolution methods to environmental problems include Higdon et al. \cite{process-convolution-Higdon99} and Calder \cite{process-convolution-Calder08}, who used the method to ground and air pollution, respectively, over large areas based on point measurements.

\section{Deformation approach}
The spatial deformation approach \cite{nonstationary-Handbook,sampson1992nonparametric} is a non-parametric method to model non-stationary and anisotropic covariance structures, which has been used , for example, in \cite{deformation-Damian03} for air pollution. The method assumes repeated samples of the stochastic field; the repeated samples are often seen as being from different times. The main idea is to relate the spatial coordinates of the sampling locations to a new set of coordinates representing a stationary spatial covariance, i.e. a non-stationary field in the original coordinates has a stationary representation in the new, or transformed, coordinates. 

Consider $Y_{it}=Y(u_i,t)$ observations at locations $u_i , \; i=1,\dotsc, N$ locations and times $t = 1, \dotsc, T$. Assuming a time constant mean the spatio-temporal process is written as
\begin{equation*}
Y(u,t)=\mu(u)+e(u,t)+\varepsilon(u,t),
\label{eq:deform}
\end{equation*}
where $\mu(u)$ represent the mean field, $e(u,t)$ is a mean zero spatio-temporal process and $\varepsilon(u,t)$ is measurement error which is independent in space and time, and independent of $e(u,t)$. 
Under these assumptions, i.e. $e(u,t)$ being non-stationary in $u$ and having independent replicates in time, one can express the spatial dispersion as
\begin{eqnarray*}
D^2(u_i,u_j)&=&V\big(Y(u_i,t)-Y(u_j,t)\big)\\
&=& V\big(e(u_i,t)-e(u_j,t)\big)+V\big(\varepsilon(u_i,t)-\varepsilon(u_j,t)\big).
\label{?}
\end{eqnarray*}
where $D^2(u_i,u_j)$ is a variogram, see Definition \ref{def:semiV} and section \ref{sec:PE}. 
All the common variogram models in geostatistical practices can be expressed as a function of Euclidean distances between site locations in a bijective transformation of the geographic coordinate system
\begin{equation}
D^2(u_i,u_j)=g\big(\vert f(u_i)-f(u_j)\vert\big)
\label{eq:dispersion}
\end{equation}
where $f$ is a transformation that expresses the spatial Non-stationarity and anisotropy, and $g$ is an appropriate monotone function, i.e. a variogram. The space of original geographic coordinates is called G-space and the space of deformed coordinates is called D-space with the deformation given by the mapping $f$, i.e. $u\in G \xrightarrow{\text{f}} f(u)\in D$. Typically, the measurements are taken in G-space two dimensional and 
\begin{equation}
f: \underbrace{\mathbb{R}^2}_{\text{G-space}}\longrightarrow \underbrace{\mathbb{R}^d}_{\text{D-space}}.
\label{eq:G-D}
\end{equation}
Deformation approach has been applied on environmental data, for instance, Damian et al. \cite{deformation-Damian03} used this method to model air pollution. Details on how to choose $g$ and $f$ are given by \cite{nonstationary-Handbook} and \cite{sampson1992nonparametric}. In the next section, one of the method of choosing $f$ that allows for covariates to affect the non-stationarity will be explained.

\section{Allowing for covariates in the covariance structure}
The main idea of this method is to consider the covariates at each location,  $u$, in the covariance structure. This leads to an increase in the dimensionality of the latent space and \eqref{eq:G-D} becomes a function from  $\mathbb{R}^2 \times \mathbb{R}^{d-2}$ to $\mathbb{R}^{d}$. This method arise in order to overcome the main problem of the deformation approach, i.e. mapping may fold so that two different points in G-space result in one point in D-space. Schmidt et al. \cite{schmidt2011considering} showed that $f$ in \eqref{eq:dispersion} can be approximated by $f(\mathbf{u}) \in \mathcal{N} (\mathbf{\mu}_f(\mathbf{u}),{\boldsymbol\Sigma}_f)$ where $\mu_f(\mathbf{u})$ is $d\times n$ matrix which contains the two coordinates of G-space and $d-2$ covariates, $\mathbf{\mu}_f(\mathbf{u})=(u_1,u_2,B_1(\mathbf{u}), \dotsc,B_{d-2}(\mathbf{u}))^T$. Clearly, the problem of folding is solved by adding $d-2$ dimension to G-space.
\\
\chapter{Handling large spatial data sets}
\label{sec:largedata}
In addition to the issue of non-stationary covariance structures so far discussed in this paper, another issue in spatial statistics is the so called \textbf{"big $N$ problem"}, or how to handle large spatial data sets. The problem can be illustrated by studying the log-likelihood, or the reconstruction at unknown sites.
Recall that the log-likelihood and the reconstruction are given by 
\begin{equation}
\ell(\theta|\mathbf{Y})= constant-\frac{1}{2}\log\vert \Sigma(\theta)\vert -\frac{1}{2}(\mathbf{Y}-\mu(\theta))^T\Sigma(\theta)^{-1}(\mathbf{Y}-\mu(\theta)),
\label{eq:loglikelihood}
\end{equation}
and
\[
X|Y, \theta \in \mathcal{N} \left( \mu_x + \Sigma_{xy}\Sigma_{yy}^{-1}(Y-\mu_x),\: \Sigma_{xx}-\Sigma_{xy}\Sigma_{yy}^{-1}\Sigma^T_{yx} \right).
\]
For $N$ observations, the computational time for $\ell(\theta|\mathbf{Y})$ scales as $\mathcal{O}(N^3)$ due to $\vert \Sigma \vert$ and $\Sigma^{-1}$ while the memory requirement for $\Sigma$ scales as $\mathcal{O}(N^2)$.

There are several computationally efficient approaches for alleviating the big N problem. Hence, a number of methods that can be divided into three main classes will be discussed in this section: \begin{inparaenum}[1\upshape)]
\item \textbf{Low rank approximation}: uses exact computations on a reduced rank or simplified version of the field, thus reducing the size of the $\Sigma$ matrices. These methods include: \begin{inparaenum}[a\upshape)]
 		\item \textbf{Fixed rank kriging} \cite{cressie2008fixed},
 		\item \textbf{Predictive process} \cite{banerjee2008gaussian} and
		\item \textbf{Process convolution} \cite{higdon1998process} which was discussed in Section \ref{ProCon} page \pageref{ProCon}.
 	\end{inparaenum}
\item \textbf{Covariance tapering} \cite{furrer2006covariance} sets small values of the covariance, $r(h)$ to zero, obtaining a sparse covariance matrix, $\Sigma$.
\item Fitting \textbf{Gaussian Markov Random Fields} (\textbf{GMRF}) to Gaussian Random Fields (\textbf{GRF}).
\end{inparaenum}
Note that both 2) and 3) modify the covariance matrix to obtain a sparse matrix which decreases both computational times and storage, see Section \ref{sec:CovTap} and \ref{sec:GMRF}. 

Large data sets are common in environmental sciences and application of the methods presented here include: Shi and Cressie \cite{fixed-ENV:ENV864} and Cressie and Johannesson \cite{fixed-cressie2008} who used fixed rank kriging, respectively on satellite measurements of ozone and aerosols; Latimer et al. \cite{predictive-process-ELE:ELE1270} used predictive process for modelling of invasive species; Furrer et al. \cite{furrer2006covariance} used covariance tapering for modelling large climatological precipitation data set; and Cameletti et al. \cite{GMRF-Cameletti12} used GMRF for modelling of particulate matter concentration.

\section{Low rank approximation}
The main idea in Low rank approximations is to express the Gaussian process $\mathbf{X(u)}$ through a set of, $r \ll n$, basis functions $\lbrace\psi_i\rbrace_{i=1}^r$
\begin{equation}
\mathbf{X(u)}= \sum_{i=1}^r \psi_i(\mathbf{u})w_i
\label{eq:basis}
\end{equation}
where $r$ is fixed and $w_i \in \mathcal{N}(0,\Sigma_w)$. 

This gives $\mathbf{X}=\mathbf{\Psi}\textbf{w} \in \mathcal{N}(\mathbf{0},\mathbf{\Psi}\mathbf{\Sigma_w}\mathbf{\Psi}^T)$, and \eqref{eq:estimation} becomes
\begin{eqnarray}
E(\mathbf{X}|\mathbf{Y})&=& \mathbf{\Sigma_{xy}}\mathbf{\Sigma_{yy}}^{-1}\mathbf{Y}= (\mathbf{\Psi}_x\mathbf{\Sigma_w}\mathbf{\Psi}^T)(\mathbf{\Psi}\mathbf{\Sigma_w}\mathbf{\Psi}^T+\mathbf{\Sigma_\varepsilon})^{-1}\mathbf{Y}
\label{eq:basisestimate}
\end{eqnarray}
where $\mathbf{\Sigma_\varepsilon} = \sigma^2\mathbb{I}$. The most expensive part of this calculation remain at $(\mathbf{\Psi}\mathbf{\Sigma_w}\mathbf{\Psi}^T+\mathbf{\Sigma}_\varepsilon)^{-1}$, using the matrix inversion lemma this simplifies to
\[
(\mathbf{\Psi}\mathbf{\Sigma_w}\mathbf{\Psi}^T+\mathbf{\Sigma}_\varepsilon)^{-1}=\mathbf{\Sigma}_\varepsilon^{-1}-\mathbf{\Sigma}_\varepsilon^{-1}\mathbf{\Psi}(\mathbf{\Psi}^T\mathbf{\Sigma}_\varepsilon^{-1}\mathbf{\Psi}+\mathbf{\Sigma_w^{-1}})^{-1}\mathbf{\Psi}^T\mathbf{\Sigma}_\varepsilon^{-1}
\]
where $(\mathbf{\Psi}^T\mathbf{\Sigma}_\varepsilon^{-1}\mathbf{\Psi}+\mathbf{\Sigma_w^{-1}})$ and $\mathbf{\Sigma_w}$ are a $r\times r$ matrices. Therefore, the computational time for the inverse matrix is $\mathcal{O}(r^3)$. Since $r \ll n$ this drastically decreases the computational time. 

There are many possible low rank methods and some of the methods are stronger in theory than in practice. The Karhunen-Lo\'{e}ve expansion uses eigenfunctions of $\Sigma$ and is only feasible for the few cases where eigenfunctions can be found analytically, (see \cite{low-Handbook}).
Here, the focus is on methods which are computationally feasible. For example, the process convolution, see section \ref{ProCon}, \eqref{eq:procon} can be seen as a low rank approximation with basis functions $\psi_i(u)=k(u-s_i)$. In the following, two other methods, fixed rank kriging and predictive processes are explained.

\subsection{Fixed rank kriging}
Fixed rank kriging \cite{cressie2008fixed} uses multiresolutional functions, often wavelets, to construct a low rank, non parametric covariance matrix. Allowing the covariance to capture variation at several scales.

In this case the covariance matrices are given by $\mathbf{\Psi}\mathbf{\Sigma_w}\mathbf{\Psi}^T$ where $\mathbf{\Sigma_w}$ is a positive definite matrix such that  $(\mathbf{\Psi}\mathbf{\Sigma_w}\mathbf{\Psi}^T+\mathbf{\Sigma_\varepsilon})$ is a close approximation of the empirical covariance estimate, $\hat{\Sigma}_{yy}$. Then   
\[
\hat{\mathbf{\Sigma}}_\mathbf{w}=\mathbf{R}^{-1}\mathbf{Q}^T(\mathbf{\hat{\Sigma}_{yy}-\Sigma_\varepsilon)}\mathbf{Q}(\mathbf{R}^{-1})^T
\]
where $\mathbf {\bar{\Psi}}=\mathbf{QR}$ is a $\mathbf{QR}$-decomposition of a matrix containing the wavelets. Strictly, $\Psi$ is a binned version since the $\hat{\Sigma}_{yy}$ estimate is computed using bins similar to those in \eqref{eq:LS-min}.

For example, \cite{cressie2008fixed} used the local bisquare function at three different resolutions,
\[
\Psi_{i(l)}(\textbf{u})= \left\{ 
\begin{array}{l l}
   \Big(1-\big(\dfrac{\Vert\textbf{u}-v_{i(l)}\Vert}{r_l}\big)^2\Big)^2 & \quad \Vert\textbf{u}-v_{i(l)}\Vert\leq r_l\\
   0 & \quad \text{o.w}
  \end{array} \right.
\]
where $v_{i(l)}$ is one of the center points of the $l$th resolution, $r_l=1.5\cdot d_l$ and $d_l$ is shortest distance between center points of the $l$th resolution. In fixed rank kriging non-parametric covariance function is specified and covariance is estimated as a non-parametric, low-rank approximation to $\hat{\Sigma}_{yy}$. 

\subsection{Predictive process}
The main idea in predictive processes \cite{banerjee2008gaussian} is similar to that of fixed rank kriging, however the difference that predictive process is based on a covariance function representation of the covariance matrix. In the predictive process the covariance function of the field is assumed to be of a parametric form, as described in Section \ref{sec:Theory} and Table \ref{covfun}. In this method, the Gaussian process in \eqref{eq:basis} is replaced with a lower ranked process $\tilde{w}$, a predictive process which is derived from the parent process $w$.
Consider a set of knots $U^*=\lbrace u_1^*,\cdots ,u_r^*\rbrace$ which may or may not form a subset of the observed locations, then the Gaussian process evaluated at the knots is $w(u^*_i)_{i=1}^r=w^*\in\mathcal{MVN}(0,\Sigma^*)$, where $\Sigma^*=cov(u^*_i,u^*_j)$ is a $r\times r$ matrix. According to \eqref{eq:basisestimate}, the reconstruction at a site $u_0$ is given by $E(w(u_0)|w^*)=\tilde{w}(u_0)=\Sigma_{u_0}^T\Sigma^{*-1}w^*$ where $\Sigma_{u_0}=cov(u_0,u_i^*)$, is the covariance between each knot and the site $u_0$. The reconstruction defines a spatial process $\tilde{w}(u)\in\mathcal{N}(0,\Sigma_{\tilde{w}})$ with covariance function $\Sigma_{\tilde{w}}=\Sigma_u^T\Sigma^{*-1}\Sigma_u$ where $\Sigma_u=cov(u,u^*_i)$. $\tilde{w}$ is called \textbf{predictive process}. It can be show that $\tilde{w}$ is the best approximation for parent process $w$, (see, \cite{banerjee2008gaussian} Section 2.3 ). The result is a process defined on only the $r$ knots that uses kriging to approximate the entire field.

\section{Covariance tapering}
\label{sec:CovTap}
Tapering is a method for approximating the covariance function of large spatial fields. The basic idea is to introduce zeros into the covariance, $r(h)$, outside of a given range, $\theta$, i.e. $r(h)=0$ if $h\geq \theta$; this results in a sparse covariance matrix $\Sigma$ and corresponding speed-up in the evaluation of \eqref{eq:loglikelihood}.
The tapered covariance is defined as
\begin{equation*}
r_{\text{tap}}(h)= r(h)T_\theta(h)
\end{equation*}
where $T_\theta(h)$ is a positive definite covariance matrix with compact support, $|h|<\theta$. Decreasing $\theta$ leads to more zeros in the covariance matrix. A basic argument is that $T_\theta(h)$ needs to be of suitable shape for the estimates to remain consistent. Furrer et al.  \cite{furrer2006covariance} listed the following possible tapering functions, see Table~\ref{tapperd}, which can be used for Mat\'{e}rn covariances.
\begin{table}[H]
\centering
 \begin{tabular}{l|l}
 Taper & $T_\theta(h)\quad \text{for} \; h\geq 0$\\[1.5ex] 
 \hline
 Spherical & $\max\lbrace (1-\dfrac{h}{\theta})^2, 0\rbrace (1+\dfrac{h}{2\theta})$\\[1.5ex]
 $\text{Wendland}_1$ & $\max\lbrace (1-\dfrac{h}{\theta})^4, 0\rbrace (1+4\dfrac{h}{\theta})$\\[1.5ex]
 $\text{Wendland}_2$ & $\max\lbrace (1-\dfrac{h}{\theta})^6, 0\rbrace (1+6\dfrac{h}{\theta}+\dfrac{35h^2}{3\theta^2})$
 \end{tabular}
 \caption{Tapperd covariance function.}
 \label{tapperd}
 \end{table} 

\section{Approximating Gaussian Random Fields with Gaussian Markov Random Fields}
\label{sec:GMRF}
First, in this section some useful definitions such as, Markov property and neighbourhood of an observation set are stated and then the section will be continued with explaining the idea of approximating GRFs with GMRFs.

For a time discrete stochastic process $x_t$, the process has a Markov property if, for any $t$, the distribution of $x_t$ given the entire history is equal to the distribution of $x_t$ given just $x_{t-1}$,
\[
p(x_t|x_{t-1},\dotsc,x_0)=p(x_t|x_{t-1}).
\]
For the spatial case, the neighbourhood of a point $u_i$ is defined as a set of neighbours $\mathcal{N}_i$, $\left\lbrace u_j,\; j \in \mathcal{N}_i\right\rbrace $  which are in some senses close to $u_i$.
Now a GMRF can be defined. A GRF, $\mathbf{z} \in \mathcal{N}(\mu,\Sigma)$ is a GMRF if the full conditional distribution for all $z_i$ satisfies
\[
p(z_i|\lbrace z_j; \; j\neq i\rbrace)= p(z_i|\lbrace z_j; \; j \in \mathcal{N}_i\rbrace)
\]
for some neighbourhood set, $\lbrace\mathcal{N}_i\rbrace_{i=1}^n$, i.e. the distribution of $z_i$ given the whole field is equal to distribution of $z_i$ given just the neighbours in neighbourhood set.

The Markov property with respect to a neighbourhood set leads to a zero-pattern in the precision matrix, $\mathbf{Q} = \Sigma^{-1}$
where 
\[
\mathbf{Q}_{ij}=0  \Longleftrightarrow  j \notin \lbrace\mathcal{N}_i, i\rbrace
\]
Using the precision matrix the density function \eqref{eq:normdensity} becomes
\[
p(\mathbf{z})= \dfrac{|\mathbf{Q}|^{1/2}}{(2\pi)^{N/2}}exp\left( -\frac{1}{2}(\mathbf{z}-\mu)^T\mathbf{Q}(\mathbf{z}-\mu)\right) .
\]
and the joint density \eqref{eq:condition} becomes
\[
\textbf{Z} =
 \begin{bmatrix}
 \textbf{X}\\
 \textbf{Y}
\end{bmatrix}
\in \mathcal{N}
\left( 
\begin{bmatrix}
 \mu_{\mathbf{x}}\\
 \mu_{\mathbf{y}}
\end{bmatrix}
,
\begin{bmatrix}
 \mathbf{Q}_{\mathbf{x}\mathbf{x}}& \mathbf{Q}_{\mathbf{x}\mathbf{y}} \\
 \mathbf{Q}_{\mathbf{y}\mathbf{x}}& \mathbf{Q}_{\mathbf{y}\mathbf{y}} 
\end{bmatrix}^{-1}
\right),
\]
with conditional distribution
\[
X|Y \in \mathcal{N} \big( \mu_x - \mathbf{Q}^{-1}_{xx}\mathbf{Q}_{xy}(Y-\mu_y),\: \mathbf{Q}^{-1}_{xx} \big).
\]
One can express the positive definite precision matrix as $\mathbf{Q} = \mathbf{R}^T\mathbf{R}$ where $\mathbf{R}$ is Cholesky factor. $\mathbf{R}$ is a unique upper triangular matrix with strictly positive diagonal elements. If $\mathbf{Q}$ is sparse $\mathbf{R}$ is often also sparse. Instead of computing $|\mathbf{Q}|$ and $\mathbf{Q}^{-1}$, one can use involving $\mathbf{R}$. The trick here is that one should never calculate $\mathbf{R}^{-1}$ (the inverse matrix), but instead solve the triangular equation system $\mathbf{Ra=b}$ to obtain $\mathbf{b=\mathbf{R}^{-1}a}$.
Using the sparsity of the precision matrix and its sparse Cholesky factorization leads to efficient computation which decreases the computational time.

To use a GMRF one needs to construct a useful and sparse $\mathbf{Q}$ matrix. A simple and common method is to use a conditional autoregressive model (\textbf{CAR}). The problem with \textbf{CAR} models is that they are restricted to lattices. Recent work by Lindgren et al.~($2011$) developed an explicit link between GMRFs and Mat\'{e}rn covariance functions. 
The method is based on the fact that a GRFs with Mat\'{e}rn covariance function on $\mathbb{R}^d$ are solutions to a Stochastic Partial Differential
Equation (\textbf{SPDE}) \cite{whittle1954stationary,whittle1963stochastic}.
\begin{equation}
(\kappa^2-\Delta)^{\frac{\alpha}{2}}Z(\mathbf{u}) = \tau\mathcal{W}(\mathbf{u})
\label{eq:SPDE}
\end{equation}
where $\mathcal{W}(\mathbf{u})$ is a Gaussian white noise, $\Delta = \nabla^T\nabla = \dfrac{\partial^2}{\partial u_x^2}+\dfrac{\partial^2}{\partial u_y^2}$ is the Laplacian and $\alpha=\nu + \frac{d}{2}$. 
Here, a simplified solution sketch for the case $\alpha = 2$ is shown. The left hand side of \eqref{eq:SPDE} can be written as $\mathbf{Kz}$ where $\mathbf{K}$ is a finite difference approximation of  $(\kappa^2-\Delta)$ and $\mathbf{z}$ is a discretized vector of $\mathbf{Z(u)}$. Hence, the discretized \textbf{SPDE} becomes
\[
\mathbf{Kz}\stackrel{d}{=} {\boldsymbol\varepsilon}
\]
where ${\boldsymbol\varepsilon}$ is a Gaussian vector with mean zero and covariance matrix $\tau^2\mathbf{C}$. Therefore, $\mathbf{z} \in \mathcal{N}(0,\tau^2\mathbf{K^{-1}CK^{-T}}) $ is a solution to \eqref{eq:SPDE} with $\mathbf{Q}=\frac{1}{\tau^2} \mathbf{K^TC^{-1}K}$. The matrix $\mathbf{K}$ can be written as $\mathbf{K}= \kappa^2 \mathbf{C}+ \mathbf{G}$ given suitable matrices $\mathbf{C}~ \text{and}~ \mathbf{G}$. Therefore $\mathbf{Q}$ is sparse if $\mathbf{K}$ and $\mathbf{C}^{-1}$ are sparse. To obtain a sparse $\mathbf{C}^{-1}$ one needs to approximate $\mathbf{C}$ with a diagonal matrix $\tilde{\mathbf{C}}$. The resulting precision matrix $\mathbf{Q} = \frac{1}{\tau^2} \mathbf{K^T\tilde{C}^{-1}K}$ is now sparse sine $\mathbf{G}$ is sparse (essentially G only contains finite difference approximation of $\Delta$).

For $\alpha=2~(\nu=1)$ the local structure of the precision matrix on a regular grid in $\mathbb{R}^2$ is given by
\[
\kappa^4h^2 \underbrace{\Bigg[\qquad 1 \qquad \Bigg]}_\mathbf{C} + 2\kappa^2 \underbrace{\left[
\begin{array}{c c c} 
  & -1 & \\
 -1 & 4 & -1\\ 
 &-1 & 
 \end{array}
 \right]}_{\approx-\Delta (\mathbf{G})}
 +\frac{1}{h^2}\underbrace{\left[
 \begin{array}{c c c c c}
  & & 1 & & \\
  & 2&-8&2&\\
  1&-8&20&-8&1\\
  & 2&-8&2&\\
  & & 1 & & \\
 \end{array}
 \right]}_{\approx \Delta^2 (\mathbf{G_2}=\mathbf{G}\mathbf{\tilde{C}}^{-1}\mathbf{G})}.
\]
In conclusion, the approximate, discretized, solution to \eqref{eq:SPDE} is $\mathbf{z}\in \mathcal{N}(0,\mathbf{Q}^{-1}_{\alpha,\kappa})$ for different integer values of $\alpha$ with 

\begin{table}[H]
\centering
\begin{tabular}{l l l} 
 $\mathbf{Q}_{1,\kappa}=\frac{1}{\tau^2}\mathbf{K}$ & $=\frac{1}{\tau^2}\left(\kappa^2\mathbf{C}+\mathbf{G}\right)$ & $\alpha = 1$\\[0.5cm]
 $\mathbf{Q}_{2,\kappa}=\frac{1}{\tau^2}\mathbf{K}\mathbf{C}^{-1}\mathbf{K}$ & $=\frac{1}{\tau^2}\big(\kappa^4\mathbf{C}+2\kappa^2\mathbf{G}+\underbrace{\mathbf{G}\mathbf{C}^{-1}\mathbf{G}}_{\mathbf{G2}}\big)$ & $\alpha = 2$\\[0.5cm]
$\mathbf{Q}_{\alpha,\kappa}=\frac{1}{\tau^2}\mathbf{K}\mathbf{C}^{-1}\mathbf{Q}_{\alpha-2,\kappa}\mathbf{C}^{-1}\mathbf{K}$ & & $\alpha = 3, 4, \dotsc$ 
\end{tabular}
\end{table}
Now the model defined in \eqref{eq:yAx} and \eqref{eq:xmunu} can be expressed using a GMRF model as
\begin{equation}
\mathbf{Y=AX}+{\boldsymbol\varepsilon} \qquad \: {\boldsymbol\varepsilon}\in \mathcal{N}(\mathbf{0,\mathbb{I}\sigma^2_{\boldsymbol\varepsilon}})
\label{eq:GMRF1}
\end{equation}
\begin{equation}
\mathbf{X}={\boldsymbol\Psi}\mathbf{w}+\mathbf{B}{\boldsymbol\beta} \qquad \mathbf{w}\in \mathcal{N}(\mathbf{0,Q^{-1}})
\label{eq:GMRF2}
\end{equation}
where \textbf{w} is a GMRF and $\mathbf{B}{\boldsymbol\beta}$ is regression specifying the mean and ${\boldsymbol\varepsilon}$ is Gaussian noise. 
\\
\chapter{Application to the LANDCLIM data}
\label{sec:application}
\section{Introductory}
Vegetation/Land-cover is an important part of the climate system with changes in vegetation abundance affecting climate and vice versa. 
Many Global Climate Models (GCMs) or Regional Climate Models (RCMs) consider vegetation as a given boundary condition, making  good estimates of vegetation is important for accurate climate reconstruction and prediction.
Recent studies, \cite{sugita2007Love,sugita2007Reveals,gaillard2008use,gaillard2010holocene} show the importance of pollen based estimates of the vegetation abundance in the reconstruction of past vegetation and land-cover.

The main purpose of the current study is to use the spatial and spatio-temporal tools explained in Section \ref{sec:one}-\ref{sec:largedata} to reconstruct vegetation and land-cover during different time windows. The reconstruction will be based on estimated vegetation proportions provided by pollen data from specific locations, see Section \ref{sec:Data} and Figure \ref{fig:reveals}. 

\section{Data}
\label{sec:Data}
The study area covers Northwest and Western Europe North of the Alps. The region has been divided into a spatial grid of $1^\circ \times 1^\circ$ $(\text{roughly}~ 111.2 \times 111.2~ \text{km}^2)$. The pollen based reconstructions are available for three time windows; two time windows in the past and the modern time; 
\begin{description}
\item[$\mathbf{6000}$ BP]\footnote{BP: Before Present}$[$\textbf{BC} $4250-3750 \approx 5700-6200$ \textbf{BP}$]$: a period with little human-induced landscape openness,
\item[$\mathbf{200}$ BP] $[$\textbf{AD} $1600-1850 \approx 100-350$ \textbf{BP}$]$: 
the classical pre-industrial state widely used as a baseline to compare modern human-impact in terms of greenhouse gases on climate change with past non-human impacted climate.
\item[$\mathbf{0}$ BP ]$[$\textbf{AD} $1850-x$\footnote{Date of the most recent available data}  $\approx x-100$ BP$]$: a modern time window which will be used to validate models.
\end{description}
\subsection{REVEALS, pollen-based vegetation reconstruction}
\begin{figure}[h!]
\centerline{\includegraphics[scale=0.7]{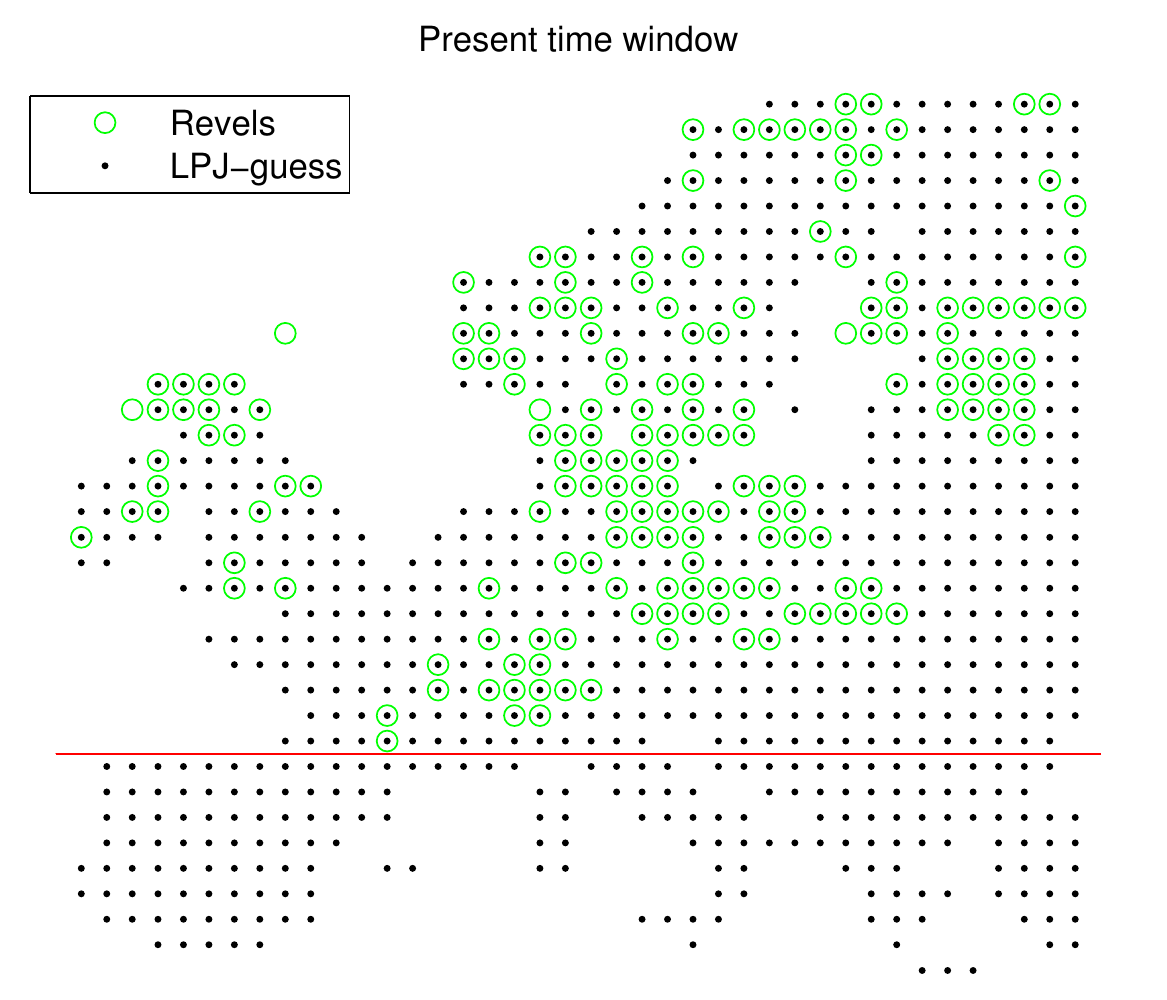} }
 \caption{The available Reveals data (174 locations) for modern time window in the study area, above the red line.}
 \label{fig:reveals} 
 \end{figure}

Estimates of vegetation are based on the Reveals model (Regional Estimates of VEgetation Abundance from Large Sites) which estimates regional vegetation/land-cover compositions based on pollen counts in sediment cores from large lakes \cite{sugita2007Reveals}.
The Reveals estimates provide abundances for $25$ taxa which can be grouped into $10$ plant functional types (PFTs) or $3$ land-cover types (LCTs), see Table \ref{tab:PFT}. The PFTs are typically used in vegetation modelling while LCTs are applied in climate modelling. Figure \ref{fig:reveals} shows the available Reveals data (174 locations) for the modern time window covering about $40\%$ of the study area.
 
\begin{table}[H]
\small

\centerline{\begin{tabular}{l l l c}
  		PFT & Description & Plant Taxa & LCT\\
   			\hline
   			\hline
   		TBE1 & Shade-tolerant-boreal & \textit{Picea} & \multirow{4}{*}{\textbf{Ever green}}\\
   		TBE2 & Shade-tolerant-temperate & \textit{Abies} & \\
   		IBE  & Shade-intolerant-boreal& \textit{Pinus} & \\
   		TSE  & Tall shrub & \textit{Juniperus} & \\
   		\hline
   		TBS  & Shade-tolerant-temperate & \textit{Carpinus, Fagus, Tilia}, & \multirow{5}{*}{\textbf{Summer green}}\\
   		&&\textit{Ulmus} &\\
   		IBS  & Shade-intolerant-boreal & \textit{Alnus, Betula, Corylus},  & \\
   		&&\textit{Fraxinus, Quercus}&\\
   		TSD  & Tall shrub & \textit{Salix}& \\
   		\hline
   		LSE  & Low evergreen shrub & \textit{Calluna vulgaris} &\multirow{6}{*}{\textbf{Open land}}\\
   		GL   & Grassland - all herbs & \textit{Artemisia, Rumex acetosa}-t & \\
   		&&\textit{Poaceae, Plantago lanceolata},&\\
   		&&Cyperaceae, \textit{Plantago montana}, &\\
   		&&\textit{Filipendula, Plantago media} &\\
   		AL   & Agricultural land- cereals & Cerealia-t, \textit{Secale}& \\
   		\end{tabular} }
   		\caption{shows the classification of $25$ taxa in $10$ PFTs and $3$ LCTs.}
 \label{tab:PFT}
\end{table}

\subsection{Additional data}
Vegetation abundance is strongly related to covariates, such as elevation, geographical coordinates, human land-use and natural potential vegetation proportions. For each time windows the natural potential vegetation proportions have been simulated using LPJ-guess (Lund-Potsdam-Jena)-(General Ecosystem Simulator). LPJ-guess is a dynamic, process based vegetation model that simulates vegetation dynamics base on climate data input \cite{smith2001representation}. Human land-use is provided in form of the  KK scenario \cite{kaplan2009prehistoric} estimates. 

In our model we use the natural vegetation from LPJ-guess adjusted for the human impact estimates from KK scenario. The adjustment is needed since LPJ-guess only models natural openness, without considerations for human impact such as farmland and pastures.
 
\section{Model}
An important feature of the Reveals data is that it is proportions of vegetation. Meaning that observations, $y_i$ have to sum to one, $\sum_i y_i =1$, and be non-negative, $y_i\geq 0$, in reality LCTs are positive. Aitchison \cite{aitchison1986statistical} modelled the compositional data by logistic normal distribution that is a multi-normal distribution via log transformation. 
The log transformation can be done in different ways depends on the data set. In our model we used central log ratio (clr) 
\begin{equation*}
u_i(s)=\log \dfrac{y_i(s)}{\sqrt{\prod}_i y_i(s)}
\end{equation*}
where $s=1,\cdots, S$ are the data locations. To ensure identifiability we have the condition $ \sum_i u_i(s) = 0$. Using $u_i$ for modelling is easier than $y_i$ due to easier constraint on $u_i$; $u_i$ can take values in $(-\infty, \infty)$ with only a sum to zero constraint. 
After using $u_i(s)$ in the model, the results are transferred back to the original space by
\begin{equation*}
y_i(s)=\dfrac{\exp(u_i(s))}{\sum_i \exp(u_i(s))}
\end{equation*}
where now $\sum_i y_i = 1, \; y_i \geq 0.$

Using the model in \eqref{eq:GMRF1} we can reconstruct the latent field using \eqref{eq:GMRF2} given observed data at specific locations. The mean value of the field can be explain using a regression model. Hence, the first task is to identify important explanatory variables. Knowing the coefficients of the regression one can reconstruct the mean field. Thereafter, additional spatial structure will be assessed and a latent field model will be used to improve the reconstruction. In the Section \ref{sec:results}, preliminary results from the regression model are shown.
In the regression model we used $u_i(s)$ as a dependant variable, as explained in Section \ref{sec:reg} and Definition \ref{def:reg}, 
\[
\mathbf{u_i = \sum_p B_p{\boldsymbol\beta}_{i,p} + e_i  \qquad e_i\in \mathcal{N}(0,\sigma^2)}
\]
where $\mathbf{B}$ contains our $\mathbf{p}$ chosen covariates.


\section{Preliminary results}
\label{sec:results}
\begin{figure}[H]
\centering 
\centerline{\includegraphics[scale=0.7]{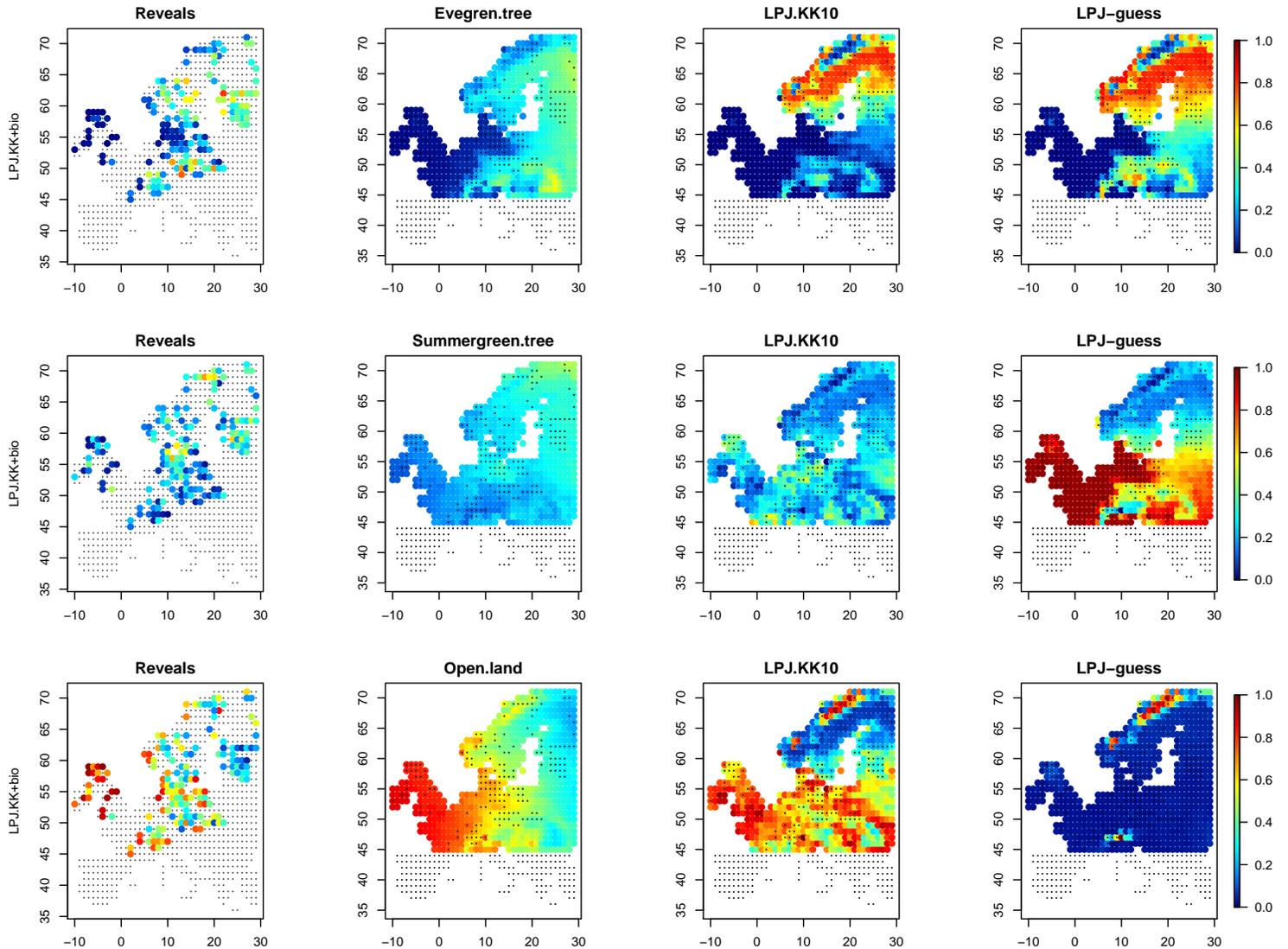}} 
 \caption{Reconstruction of REVEALS data in modern time window. First column is the original REVEALS data. Second column is the reconstruction using regression model. Third column is the adjusted LPJ-guess with KK scenario and the forth column is the natural potential vegetation from LPJ-guess.}
 \label{fig:ResPres} 
\end{figure}

The preliminary results from the regression models are available, and Figure \ref{fig:ResPres} shows the reconstruction of three different LCTs based on Reveals data set for the modern time. 
In this model, we used the adjusted LPJ-guess proportions, coordinates and elevation as covariates. 

The result shows a good fit of the model to REVEALS data and at the same time keeping the structures of the adjusted LPJ-guess mostly where we have no information from REVEALS, i.e. locations without REVEALS data.

\section{Future work}
The preliminary results show a good fit of the model to the REVEALS data set. We plan to continue the study in four main directions: \begin{inparaenum}[1)\upshape] \item identifying and evaluating covariates, \item assessing spatial structure in the data, \item attempting to identify deviations from the existing human land-use models, and \item incorporating the REVEALS reconstructions uncertainties in the model. 
\end{inparaenum}

\subsection{Identifying and evaluating covariates}
As mentioned earlier, the study contains three contrasting time windows in terms of climate and anthropogenic land-cover change, e.g. 6000 BP is considered as a relatively warm climate with low human impact. For each different time windows we will assess the main covariate and their corresponding coefficients to construct two regression models. Comparisons of the (different) models obtained for each time window will allow us to improve our understanding of land-cover and vegetation changes. In addition the results from modern time will be used to evaluate models.

\subsection{Assessing spatial structure}
Assessing the spatial structures and adding a spatial dependency structure to the model to improve the reconstruction. The preliminary results from model 2 show the effects of coordinates on the REVEALS reconstruction. This can be interpreted as the existence of spatial dependencies in the data. To capture these dependencies we plan to add a spatial field to our regression model. Expanding the model from a linear regression to 
\[
\left[ \begin{array}{c}
u_1\\
u_2
\end{array} \right]=
\left[ \begin{array}{c}
X_1\\
X_2
\end{array} \right]+\mathbf{B}{\boldsymbol\beta},
\]
where
\begin{equation}
\left[ \begin{array}{c}
X_1\\
X_2
\end{array} \right] \in \mathcal{N} \big( 0, 
\begin{bmatrix}
1&\rho\\
\rho&1
\end{bmatrix}
\otimes \mathbf{Q}^{-1}
\big).
\label{eq:field}
\end{equation}
In \eqref{eq:field}, $\mathbf{Q}$ is a sparse precision matrix the spatial structure in the field of transformed compositional data and $\rho$ is the correlation between the fields. We hope that adding spatial dependencies will improve the reconstruction \cite{tjelmeland2003composition}.

\subsection{Identifying an independent human land-use model}
The LANDCLIM project goals include quantification of human-induced changes on regional land-cover. Identifying the human impact allows for a separation of the historical process into: \begin{inparaenum}[i\upshape)]\item climate-driven changes in vegetation and \item human-induced changes in land-cover.\end{inparaenum}~We will try to identify deviations from existing human land-use models (e.g. KK scenario and HYDE) and use the deviations to construct an improved model for human impact.

\subsection{Incorporating the uncertainty}
In addition to the reconstructed vegetation composition, the REVEALS data has varying uncertainty between different locations. Right now, we are not using these available uncertainties information, this could be included, possibly using a Dirichlet observation model as in \cite{paciorek2009mapping}. Dirichlet multinomial model is also known as compound multinomial distribution which can account for this varying uncertainty in REVEALS data. It also allows for zero proportions, the issue that will be arisen if using PFTs. 
\\


 \bibliographystyle{ieeetr}
 \bibliography{introductory_bib}

\printindex
\end{document}